# Resilient Design in Nuclear Energy: Critical Lessons from a Cross-Disciplinary Review of the Fukushima Dai-ichi Nuclear Accident

Ali Ayoub[a], Haruko M. Wainwright[a], Giovanni Sansavini[b], Randall Gauntt[c], and Kimiaki Saito[d]

[a] MIT, Massachusetts, USA, aliayoub@mit.edu,
[b] ETH Zurich, Switzerland
[c] Sandia National Laboratories - retired, USA
[d] Japan Atomic Energy Agency, Japan

**Nuclear energy has been gaining momentum recently as one of the solutions to tackle climate change. However, significant environmental and health-risk concerns remain associated with potential accidents. Despite significant preventive efforts, we must acknowledge that accidents may happen and, therefore, develop strategies and technologies for mitigating their consequences. In this paper, we review the Fukushima Dai-ichi Nuclear Power Plant accident, synthesize the time series and accident progressions across relevant disciplines, including in-plant physics and engineering systems, operators' actions, emergency responses, meteorology, radionuclide release and transport, land contamination, and health impacts. In light of the latest observations and simulation studies, we identify three key factors that exacerbated the consequences of the accident: (1) the failure of Unit 2 containment venting, (2) the insufficient integration of radiation measurements and meteorology data in the evacuation strategy, and (3) the limited risk assessment and emergency preparedness. We propose new research and development directions to improve the resilience of nuclear power plants, including (1) meteorology-informed proactive venting, (2) machine learning-enabled adaptive evacuation zones, and (3) comprehensive risk-informed emergency planning while leveraging the experience from responses to other disasters.**

## Introduction

Resilience is defined as the ability to prepare for, absorb, and recover from disasters [1]. Building a resilient framework starts from the assumption that accidents—particularly low-probability but high-consequence events—are unavoidable, and then focuses on mitigating their consequences. Nuclear energy communities have made tremendous efforts to prevent accidents through advanced engineering solutions, risk assessments, and safety upgrades [2, 3]. As a result, the nuclear power industry has one of the lowest accident rates among energy industries [4]. However, accidents are possible [5-7], and several high-profile nuclear accidents have led to significant impacts on the environment and society. Currently, safety concerns still hinder the license renewal of existing nuclear power plants (NPPs) and the siting of new plants, even while there is renewed interest in nuclear power as a low-emission source.

In particular, the accident at the Fukushima Dai-ichi Nuclear Power Plant (FDNPP) on March 11, 2011, has had a significant impact not only on the local communities, but also on the choice of electricity-generation technologies across the world [8, 9]. Even though no public health consequences have been reported as a result of direct radiation exposure [10], the accident has had a negative impact on the general public health, because of prolonged relocations and emotional distress, as well as harm to the local economy. Although the FDNPP accident was initiated by an extreme natural event, the chances of other severe floodings are non-negligible and similar near-miss incidents have been documented at other locations [6]. Additionally, the extended loss of off-site power has been reported at the Zaporizhzhya Nuclear Power Plant (Ukraine) due to the Russia-Ukraine conflict in September and November 2022.

The FDNPP accident offers important lessons for improving the resilience of nuclear energy. Over the past ten years, there have been numerous studies across disciplines, analyzing the cause and progression of the FDNPP accident, and identifying mistakes and lessons for severe accident management and design improvements. One group of studies addresses the onsite engineering and management issues that led to the accident and exacerbated its consequence, identifying critical strategies for enhancing the safety of nuclear installations [11-14]. Other studies focus on the offsite emergency responses flaws, extracting lessons in emergency preparedness and radiation protection [15-18]. At the same time, many studies address the environmental and health consequences of the accident [19], estimating the amount of released radioactivity (source term) [20-22], investigating the radionuclide atmospheric transport [23, 24], and determining the long-term contamination extent [25-27].

However, there are only a few studies that cover the FDNPP accident across disciplinary boundaries, including in-plant physics and engineering systems, operators' actions, emergency responses, radionuclide release and transport, and environmental and health consequences. Although there are large government reports available [28, 29], each discipline has been often discussed separately. In particular, there have been no significant studies to investigate which particular actions or factors were the most influential on the long-term environmental and health consequences. At the same time, while some attempts have been made to extract resilience lessons following the accident, these have either been limited to the theoretical foundations [30], or have been focused on radiological resilience through radiation monitoring and mapping [31]. In addition, there have been limited discussion on how we can transfer resilience lessons from other natural and anthropogenic disasters [32, 33] into the nuclear context.

The objective of this study is to provide a comprehensive retrospective analysis of the FDNPP accident across disciplines, and to identify key gaps and potential innovations for improving resilience, particularly to mitigate the consequences of severe accidents. To make an explicit connection between onsite and offsite decision making, we reconstruct the accident timeline across different domains, namely, core and containment status, radioactive releases, and governmental protective measures, and across temporal scales, from the short minutes/hourly events up to the decade-worth long-term and future impacts. In parallel, we summarize the environmental and health consequences in areas within and outside the evacuation zones. We then discuss the key factors in the on-site/off-site emergency responses that led to significant environmental consequences. Finally, we outline the path forward to establish resilience in nuclear power based on our analysis, as well as on recent advances in broader research communities.

**Plant Design and Safety Features**

The FDNPP had six boiling water reactor (BWR) units. The reactor pressure vessel (RPV), which holds the reactor core and coolant, is designed to withstand high temperatures and pressures. The RPV is housed in the primary containment vessel (PCV), which is designed to condense steam, confine fission products in case of an accident, and provide a heat sink and water source for some safety-related systems (Fig. 1). In case of an accident or operational disruption, the reactor is shut down by inserting control rods. Different emergency core cooling systems (ECCS) are in place to remove the decay heat from the core following shutdown and prevent core melting. In the event of loss of off-site power supply, Unit 1 had an isolation condenser (IC) that uses dedicated heat exchangers to maintain core cooling through natural circulation. Units 2-5 had a high-pressure coolant injection system (HPCI) that uses the steam from the reactor to drive its pump and maintain the RPV water inventory and core cooling during loss of coolant accidents. In addition, Units 2-6 were equipped with a steam-driven reactor core isolation cooling system (RCIC), which provides makeup water from the condensate storage tank (CST) to the RPV. When the CST is depleted, the RCIC suction is switched to the suppression pool (SP), which itself acts as a heat sink for the steam generated in the RPV (steam running the turbine), hence forming a closed-loop cooling cycle.

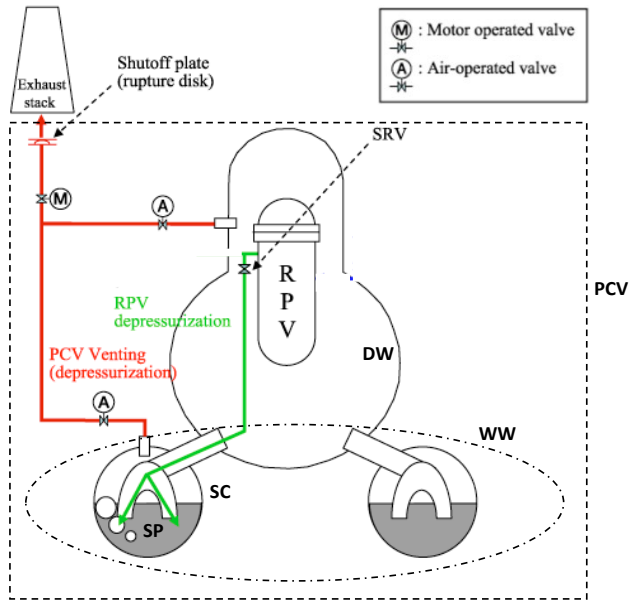

**Fig. 1.** FDNPP generic containment layout, retrieved and modified from TEPCO's schematic [34]. The safety relief valves (SRV) can be opened to discharge the RPV excess pressure into the PCV SP (green arrow) and allow the RPV low-pressure injection systems to operate. The PCV consists of an inverted bulb-shaped drywell (DW) made of 30 mm steel, backed by reinforced concrete, and connected to a torus-shaped wetwell (WW) containing the suppression pool beneath. The PCV has two redundant venting lines (red arrows): wet venting from the suppression chamber (SC) (which scrubs the radionuclides before the release to the environment), and dry venting directly from the DW. The PCV vent lines include a rupture disk that bursts when the PCV pressure exceeds the PCV design pressure.

When the ECCS are unavailable, the core temperature increases to a point at which the oxidation of the zirconium fuel cladding produces hydrogen, which, along with the generated steam during the boil off in the core, can over-pressurize the RPV. If the cladding fails, volatile fission products are released from the core. Although the PCV and the RPV are designed to withstand heavy loads (up to 0.5 MPa [35] and 8.7 MPa [28], respectively), venting is a pressure and heat release mechanism necessary to protect their integrity and to allow the low-pressure ECCS to inject water into the reactor.

**Accident Timeline and Consequences**

In this section, we reconstruct the accident timeline across different domains based on multiple reports and we connect the onsite events and actions during the FDNPP accident to their offsite consequences.

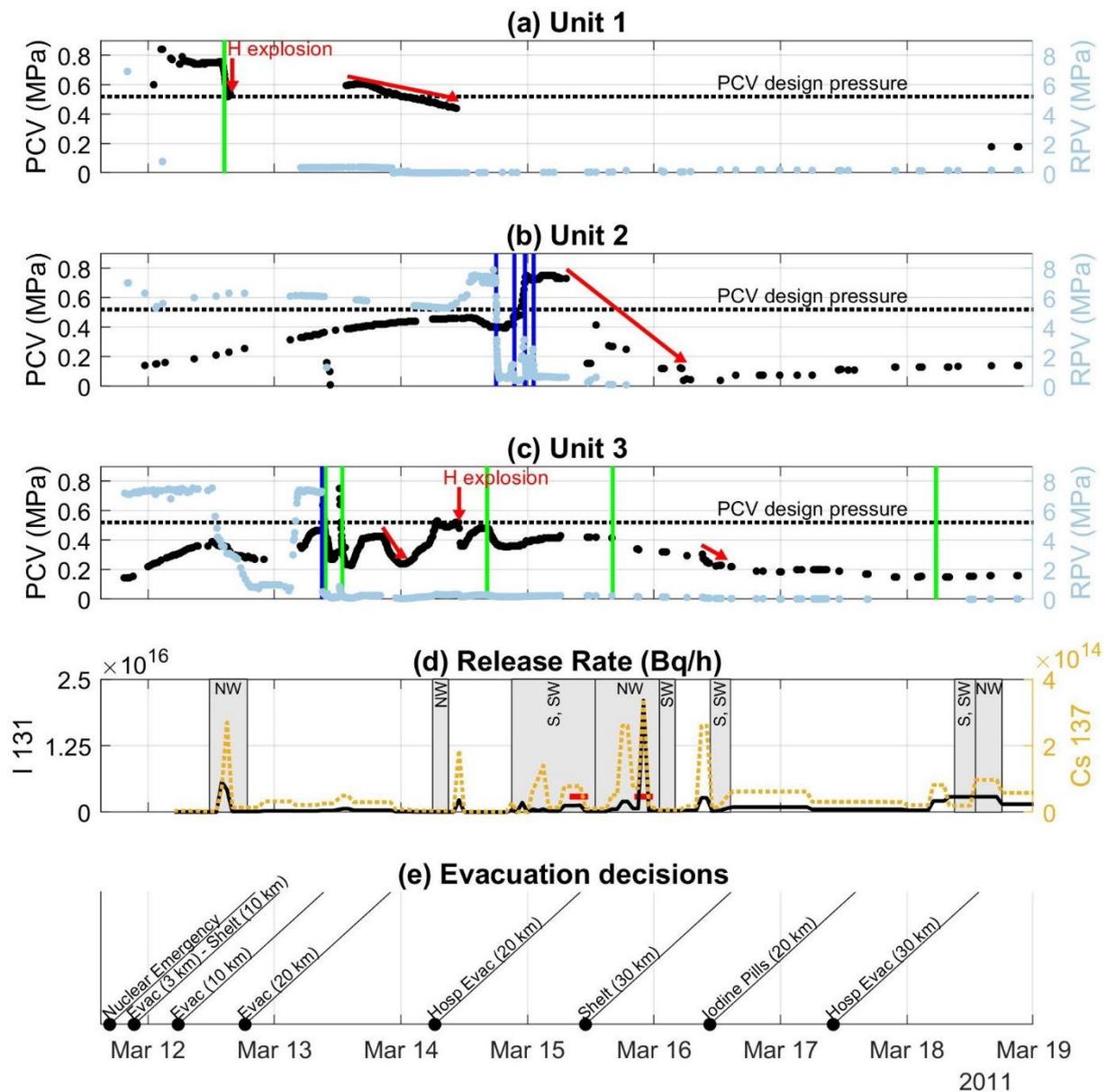

**Fig. 2.** Time sequence of the FDNPP accident: (a-c) the pressure history in the RPV (cyan) and PCV (black) of (a) Unit 1, (b) Unit 2, and (c) Unit 3; (d) release rates of $^{131}$I (black) and $^{137}$Cs (orange) radioisotopes corrected at shutdown time; and (e) emergency response decisions. In (a)-(c), the vertical blue lines mark the timings of the RPV venting events, the vertical green lines mark the timings of the successful PCV venting events or the ones with associated radioactive release, and the red arrows mark the notable events (hydrogen explosions and structural/functional PCV failures). In (d) the gray boxes represent the time windows when the prevailing wind was blowing inland, and the symbols depicted on top detail the cardinal and ordinal directions to which the wind was blowing (the direction of radionuclide spreading from the FDNPP); the horizontal red dashes cover the time intervals when TEPCO's nearby car recorded very high dose rates. Contrary to most studies, especially those focusing on the source term estimation and contamination [20, 36-38], which visualize the release rates in log scale, we present them on a linear scale to represent the actual contamination and highlight the pulse releases that happened during the accident. Note that following the tsunami and the site flooding, the monitoring of plant parameters was interrupted, then restored in the late evening of March 11.

**March 11**. The accident started with an earthquake on March 11 at 14:46 JST, causing the loss of offsite power for all units. The reactors automatically shut down, and cooling systems started operating in all the units immediately after. Unit 1 was experiencing excessive cooling rates; thus, the operators decided to stop the IC by closing its control valves at 15:03 to minimize the thermal stress on the RPV. At around 15:40, a 14 m high tsunami arrived and flooded the site, which disabled the onsite emergency power systems and caused a station blackout across all Units. Subsequent attempts to open the control valves and restart the IC failed; hence, Unit 1 had no cooling since 15:03.

The operators notified the government at 16:45 that a nuclear emergency was taking place, as specified under the Japanese Nuclear Emergency Act [39, 40]. Accordingly, the local government ordered the evacuation of residents within a 2 km radius from the plant at 20:50 (Fig. 2e). Just half an hour later, the national government issued an evacuation order for residents within 3 km of the plant and ordered sheltering within 3-10 km from the plant [15].

**March 12**. Due to the loss of cooling, Unit 1 was the first to experience core damage. Molten core material breached the RPV, which depressurized from 6.9 MPa to less than 1 MPa between the evening of March 11, and 02:45 on March 12 (Fig. 2a). The pressure in Unit-1 RPV remained low throughout the event, confirming the RPV breach and suggesting that some molten core material relocated to the PCV [41]. Concurrently, the Unit-1 PCV pressure readings started showing beyond-design values reaching 0.85 MPa at 02:45, as a result of hydrogen accumulation and gas generation from the molten core-concrete interaction [42]. After several hours of struggles to open the vent lines due to high radiation levels and limited lighting, the operator was able to vent Unit-1 PCV at 14:30 [43], which led to the recorded PCV pressure drop (Fig. 2a). At 15:36, a hydrogen explosion took place outside the Unit-1 PCV, destroying its reactor building (Fig. 3). Following the venting and the explosion, the onsite monitors detected the first significant release to the environment [28]. The estimated release rates were about $5.7 \times 10^{15}$ Bq/h for $^{131}$I and $2.7 \times 10^{14}$ Bq/h for $^{137}$Cs, which happened when the wind was blowing to the northwest (Fig. 2d). The released radioactivity is attributed to the fission products that had leaked through the PCV head and accumulated in the reactor building as a result of excessive strain that stretched the PCV head bolts [44].

In Unit 2, although the RCIC was running to keep the core cooled from the onset of the event, its water suction source (CST) was being depleted. Accordingly, the suction was switched to the suppression pool cooling mode on March 12 at 05:00 [29]. This closed-loop RCIC operation, as well as the cycling of the safety relief valves (SRVs) to keep the RPV pressure low, led to an increase in the pressure and temperature of Unit-2 PCV from March 12 onwards (Fig. 2b).

Similarly, the RCIC of Unit 3 was switched to the suppression pool cooling mode after some time, while cycling the SRVs to keep the RPV pressure low, which led to a PCV pressure increase, reaching 0.4 MPa on March 12 at 12:45 (Fig. 2c). The operators then used the fire pumps to cool the PCV, and the pressure dropped to about 0.3 MPa by 20:40. The RCIC stopped operating around noon and could not be recovered afterwards; the HPCI system started automatically [29] and brought the RPV pressure down to 1 MPa by 20:40 (Fig. 2c).

Because of the worsening situation, the national government extended the evacuation boundary to 10 km at 05:44 (Fig. 1e) before prompting Unit 1's PCV venting, which was not allowed until the evacuation was completed [28]. The wind was blowing to the ocean throughout the whole morning of March 12, but when the venting took place, the wind had shifted towards the inland. Unfortunately, the operators did not consider the wind directions when timing the venting. Following the hydrogen explosion and the associated release, the national government extended the evacuation order to 20 km at 18:25 (Fig. 2e).

**March 13**. At 13:37, the pressure measurements in Unit-1 PCV were re-established. The readings showed a continuous decrease in the PCV pressure until the end of the event (Fig. 2a) without any recorded operator actions. Such a continuous decline suggests a PCV "functional failure," in which the PCV experiences beyond-design leakage rates (see PCV failure modes in Methods).

The RCIC in Unit 2 continued to operate in the recirculation mode throughout March 13; accordingly, the PCV pressure continued to rise (Fig. 2b). Although the PCV vent valves were opened, venting could not take place because the pressure stayed below the burst pressure of the rupture disk.

Concerned about the reliability of Unit 3 HPCI turbine[1], the operators shut down the HPCI on March 13 at 02:42 and intended to use the low-pressure fire pumps—which were being used for PCV cooling—to cool the RPV. However, the attempts to open the SRVs to maintain a low RPV pressure failed, and the fire pumps did not have enough pressure head to inject water into the high-pressure RPV. With the failure to restart the HPCI, and the ensuing loss of core cooling, the RPV pressure started to increase and reached more than 7 MPa by 04:30 (Fig. 2c). Meanwhile, the PCV pressure was building up after switching the fire pumps from PCV cooling and aligning them to cool the RPV. Around 09:00, the SRVs were opened using car batteries [28]. Consequently, the RPV pressure decreased to about 0.6 MPa at 09:08, which enabled the fire pumps to inject water into the RPV [12]. However, it is estimated that the core was uncovered and had started to melt by that time [45]. With the heat deposited from the RPV, the PCV pressure reached 0.6 MPa by 09:20 (Fig. 2c), bursting the rupture disk and releasing the excess pressure through the vent lines that had been previously opened. The PCV vent valves closed several times throughout March 13, due to the "loss of air supply", and were reopened using mobile compressors (and sometimes manually). This explains the PCV pressure fluctuations in the afternoon of March 13 (Fig. 2c), and the radioactive releases associated with the venting events (Fig. 2d). The PCV pressure decrease in the late night of March 13 was attributed to a leakage (PCV functional failure) rather than a triggered venting [46], and was not associated with a significant release (Fig. 2d).

**March 14**. At around 12:30, the Unit-2 RPV pressure started to increase, implying that the RCIC, and hence the core cooling, had stopped. The operators depressurized the RPV down to 0.6 MPa by 18:47 (Fig. 2b) to allow the low-pressure fire pumps to cool the core. However, depressurizing the RPV further increased the PCV pressure, and at that time, the PCV vent valves were dislodged by the hydrogen explosion at Unit 3 [28]. Although the low-pressure injection to the RPV started around 20:00, the core was uncovered by that time and was damaged during the night of March 14 [41]. The operators continued to open more SRVs to depressurize the RPV over the following hours, and the RPV pressure was responding accordingly, i.e., rising due to the core heat-up, steam generation, and fuel melt, and dropping with the successive RPV venting operations as shown in Fig. 2b. The successive SRV openings could be associated with the small releases between 22:00 March 14 and 03:00 March 15 (Fig. 2d). Although there was no direct connection between the RPV and the environment, the over-pressurization of the PCV (reaching ~0.75 MPa by 23:40) could have resulted in the leakage of hydrogen and fission products from the PCV to the reactor building [47, 48].

Following the core melt in Unit 3, it is estimated that the RPV was breached in the early morning of March 14 [49], thus its pressure remained low afterwards, and the PCV pressure started to rise (Fig. 2c) due to the large amount of hydrogen. Similar to Unit 2, it is expected that hydrogen and some fission products escaped to the reactor building as a result of the PCV leakage [47]. The hydrogen mixed with air in the reactor building, forming an explosive gas that ignited and destroyed the reactor building of Unit 3 at 11:00 (Fig. 3). The resulting release rates were estimated at $2.3 \times 10^{15}$ Bq/h $^{131}$I and $1.8 \times 10^{14}$ Bq/h $^{137}$Cs (Fig. 2d). It should be noted that this estimate assumes that the release rate was similar to that of Unit 1 due to the lack of monitoring data at that time.

By March 14, according to GPS-enabled mobile phone logs [50], the evacuation within the 20 km zone was completed. The government ordered the emergency evacuation of the patients who were not yet evacuated from the 20 km evacuation zone. Patient evacuation started in the early morning of March 14 (Fig. 2e). Since no previous arrangements were in place, many of the patients were transported more than 200 km, with limited accompanying medical staff, to find admitting facilities [15].

**March 15**. With the sustained high PCV pressure and the failure of all venting attempts—including attempts at unfiltered venting from the drywell—Unit-2 PCV eventually failed, and its pressure dropped sharply between 07:20 and 11:20 on March 15 (Fig. 2b), reaching atmospheric pressure by the early morning of March 16. The abrupt pressure drop and the concurrent loud sound heard from the torus room [28] suggest that the Unit-2 PCV experienced a structural failure (see PCV failure modes in Methods). As a result of the March 12 hydrogen explosion in Unit 1, the reactor building of Unit 2 lost its blowout panel [28], creating an opening that enabled the accumulated hydrogen to escape and preventing its explosion [51], yet opened a direct path to the environment. It is worth mentioning that there was significant confusion at this time because Unit 4 experienced a hydrogen explosion almost at the same time as the Unit-2 PCV failure and the associated loud sound. It was later confirmed that hydrogen migrated from Unit 3 to Unit 4 through the shared stack.

---

[1] The operators were concerned about a potential HPCI turbine failure that could release contaminated steam outside of the PCV when operating on low reactor steam pressure.

The highest dose rates recorded onsite during the accident were around this period [52], specifically ranging between 8-12 mSv/h from 08:30-10:30 of March 15, and 4-8 mSv/h in the late night. Similarly, the highest estimated release rates occurred during this same period, reaching $2.1 \times 10^{16}$ Bq/h $^{131}$I and $3.4 \times 10^{14}$ Bq/h $^{137}$Cs (Fig. 2d). It is worth highlighting that these few hours of high release are estimated to sum up (by integration, see Fig. 2d) to about 31 PBq and 1.5 PBq of $^{131}$I and $^{137}$Cs, respectively. Therefore, at least 25% and 15% of the total $^{131}$I and $^{137}$Cs respective estimated releases can be attributed to the unfiltered release from the drywell of the failed Unit-2 PCV. From March 15 onwards, the Unit-2 RPV pressure remained low, suggesting that the RPV breached during that period.

In Unit 3, the operator records [49] suggest that a wetwell PCV venting took place on March 15 at 16:05, which is confirmed by the simultaneous PCV pressure drop (Fig. 2c). However, we expect that the associated radioactive release was small because most of the radioactivity was scrubbed in the wetwell (particulate radioactive material is reduced by a factor of about 1000 [29]). This can also be confirmed by the records of the onsite radiation monitors that showed no significant increase in dose rates following that venting [52].

With the alarming radiation conditions on site, the national government extended the protective actions and ordered sheltering for residents within 20-30 km from the plant on March 15 at 11:00 (Fig. 2e). After this order, the transportation of food and other supplies decreased significantly into the region, due to delivery staff concerns about radiation exposure [53].

**March 16 – onwards**. Keeping the vent valves of Unit-3 PCV open was a difficult task throughout the event, owing to the disruptions in compressed air supply. Thus, there were several recorded attempts to reopen the valves and vent the containment [49]. For example, the release-rate jumps on March 18 starting around 05:30 (Fig. 2d) coincide with a venting attempt, although no PCV pressure decrease was recorded (Fig. 2c). Conversely, the high release on March 16 between 09:00-11:00 and the continuous release afterwards do not match any operator-triggered events; however, they coincide with a pressure decrease (leakage) in Unit-3 PCV (Fig. 2c), and potentially further releases from the failed Unit-2 PCV.

The spent-fuel pool in Unit 4, which is outside the PCV, was a major concern at this time, especially after the hydrogen explosion that was initially thought to be caused by the oxidation of fuel cladding in the spent-fuel pool [47]. Therefore, the US recommended the evacuation of the US citizens living within 50-mile from the plant [47]. This discrepancy in the evacuation zone between the US and Japan added significant confusion among the residents. Later, a helicopter mission confirmed that the spent-fuel pool still contained enough water [47].

After March 21, no major releases or spikes were recorded onsite [54], and by March 22, offsite power was restored. Between March 23–25, RPV water injection was switched from sea water to borated fresh water [35]. By December 16, a cold shutdown state was reached, and the plant entered the long-term recovery phase [28].

The distribution of potassium iodide (KI) pills did not start until the morning of March 16, after many ongoing discussions within the national government regarding their side effects. However, evacuations were already completed by that time, and some evacuees had already been exposed to radioactive iodine [55].

The sheltering order remained in place until April 22, when the criteria for deliberate evacuation and relocation were established [28], and the actual public relocation started on May 15. During the extended sheltering period, daily necessities started to run out in the area, and some hospitals had to evacuate their patients (already on March 17) due to the shortage of medical supplies.

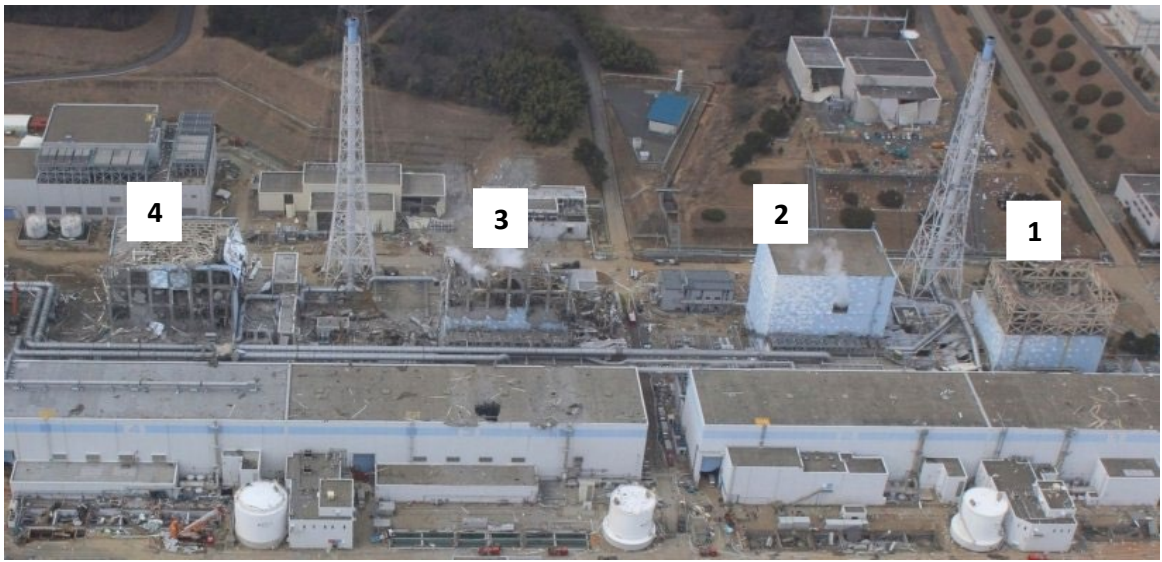

**Fig. 3.** The status of the FDNPP Units 1, 2, 3, and 4 after the accident, retrieved and modified from the American Nuclear Society schematic [56]. The hydrogen explosions destroyed the reactor buildings of Units 1, 3, and 4. Unit-4 reactor building experienced a hydrogen explosion due to the backflow of hydrogen from Unit 3 via the interconnected ventilation system [28].

**Contamination and long-term evacuation zone.** The highest radiation air dose rates during the accident were recorded in the nearby southern and southwestern regions of the FDNPP, as well as along the extended northwestern regions (Fig. 4a). These high-dose regions were along the directions of the prevailing winds during the major releases, specifically, the March 15 release from Unit 2 (Fig. 2d). In addition, precipitation (rain, fog, etc.) caused wet deposition of radionuclides in this region [57].

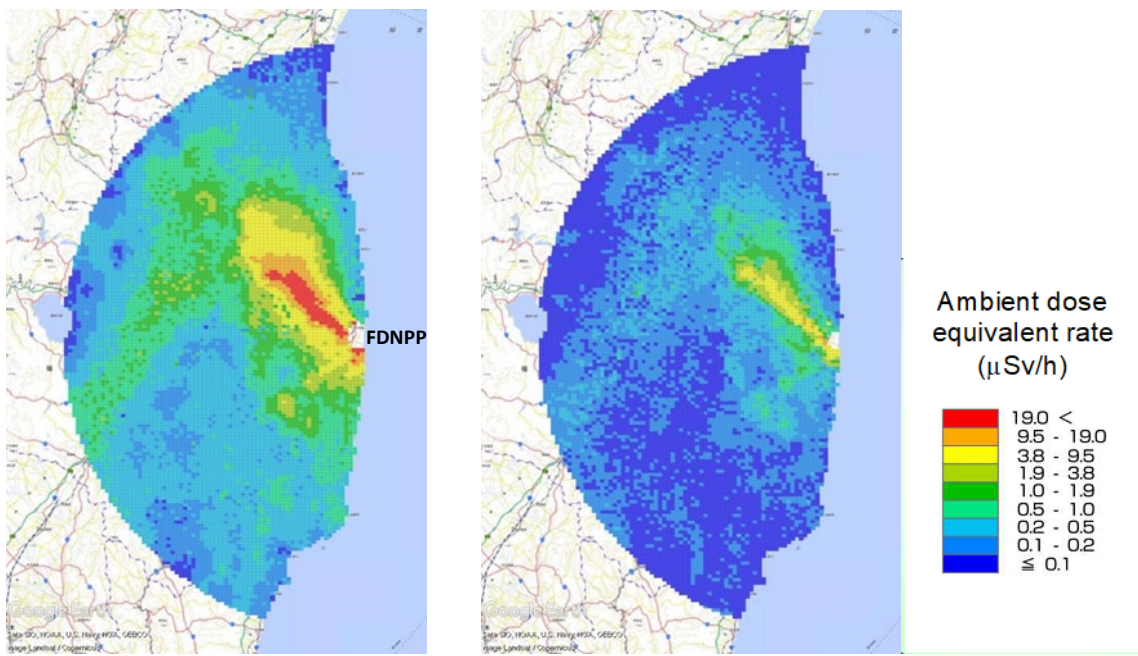

(a)                                                              (b)

**Fig. 4.** Integrated air dose rate maps in the 80 km zone in (a) 2011 and (b) 2016 [25]. The high dose rates are in the northwestern regions of the FDNPP, which is persistent over the time.

For the long-term recovery effort, the radionuclide contamination has been mapped extensively over the years to document the decline in radiation levels and support the return of evacuees [25, 26]. The main contaminant is $^{137}$Cs and $^{134}$Cs in terms of the exposure dose, after $^{131}$I decayed out (8-day half-life). Although the impacted area has been decreasing in size (Fig. 4b) following the reduction in radiation levels due to various factors, including decontamination, some parts of the northwestern regions are still designated as difficult-to-return-to zones 11 years after the accident.

**Radiation exposure and health consequences.** Investigations by international organizations report that there has been no direct deaths among residents due to radiation exposure from the FDNPP accident [10]. The Fukushima Prefecture estimates that 99.3 % of the evacuees received less than 3 mSv external dose [58, 59]. In contrast, the evacuation was delayed until May in the northwestern region outside the 20 km evacuation zone, and in some areas of this region, the cumulative exposure was later estimated to have reached 50 mSv by the time the official evacuations started [15, 60].

Several thousands who fled from the disaster zone have reported mental health issues due to family separations, isolations, social distress, and the fear of radioactivity [61]. After the accident, around 15% of the evacuees reported mental health problems and 70% reported sleeping difficulties [62]. Additionally, some studies reported a statistically significant increase in diabetes among evacuees [63]. Furthermore, according to the National Diet of Japan [64], more than 60 hospital patients and nursing-care residents in the 20 km zone died during or after evacuation due to deterioration of their medical conditions as a result of the disorganized evacuation. Studies that followed-up on evacuated patients found that they experienced mortality rates significantly higher than the average [65].

## Key Factors That Exacerbated Accident Consequences

The analysis of the FDNPP accident timeline allows to identify three key factors that have significantly impacted the accident consequences. Here, we discuss these factors in conjunction with international standards, practices, and developments.

**Containment venting.** Our analysis suggests that containment venting was the most important factor in the onsite accident management. The failure of Unit-2 PCV (March 15) caused the largest radioactive release and land contamination during the entire accident (Fig. 2d). In fact, based on the latest contamination evaluations [66], we estimate that the few hours of unfiltered release from the drywell, triggered by Unit-2 PCV failure, contributed to most of the land contamination and the subsequent 10+ years of decontamination efforts and costs.

Although there are previous studies that have suggested that Unit 2 had the major share of the radioactive release [67, 68], this study synthesized the in-plant dynamics, wind direction and source-term estimation as well as recent simulation studies to confirm the significant contribution of Unit 2, and its PCV venting failure, to the environmental consequences. In fact, most of the radionuclides from Unit 1 and 3 were considered to be scrubbed in the wet wells [42, 45], minimizing the release to the environment (Fig. 2d). This is despite the fact that Unit 1 was the first to experience core damage and a RPV breach, and that water failed to reach its core, thus leaving it without cooling for at least 11 days according to the limited available data and simulation studies [42]. Moreover, the contamination associated with the Unit 1 releases is within the 15 km radius, and shows significantly lower air dose rates than the Unit 2 plume [69]. The wetwell venting of Unit-1 and Unit-3 PCVs protected their integrity in such a way that the PCVs functioned as designed.

The prevailing philosophy in the nuclear industry was to postpone the PCV venting as long as possible in order to delay the radioactive release to the environment [28]. Furthermore, the PCV vent line design of the FDNPP, and similarly some other BWRs around the world [70], included a rupture disk that prevents any venting attempts as long as the PCV pressure is below the design pressure. Following the FDNPP accident, many countries updated their severe accident management guidelines (SAMGs) to include "early venting" as a means to preserve the containment's integrity [71, 72].

To the authors' knowledge, however, existing SAMGs around the world do not consider the prevailing wind directions before initiating PCV venting. In fact, Unit 1 was vented on March 12 at 14:30 when the wind was inland, although it was blowing toward the ocean throughout the period before (Fig. 2d), when the Unit-1 PCV pressure was already at its highest (Fig. 2a). Nevertheless, despite the favorable wind directions, the operators postponed the decision to vent until the evacuation was completed [28], and the actual venting was further delayed due to technical difficulties.

Moreover, the FDNPP, like many NPPs around the world, had no filtered containment venting system (FCVS) in place. According to Bal et al. [73], current FCVS technologies allow the retention of up to 99.9% of radioactive aerosols. Following the FDNPP accident, Japan decided to install FCVS similar to what most European NPPs have done after Chernobyl [74]. In turn, the US Nuclear Regulatory Commission still considers that FCVS are not necessary [75] and argues that venting from the wetwell provides enough filtration.

**Evacuation zone.** Owing to the defense-in-depth and the layered confinement structures in a NPP, it usually takes about 12-24 hours before a major radioactive release (Fig. 2d). During the FDNPP accident, a series of evacuation orders enabled most of the people in the 20 km radius to evacuate before March 15, when the major releases happened (Fig. 2e). However, many unexpected events happened during the accident, such as the hydrogen explosions, which complicated the situation onsite and disabled the prepared vent lines. Moreover, the accident development was uncertain and could have been worse, for example, with the potential loss of water in the Unit-4 spent-fuel pool. Given that there was substantial uncertainty regarding the plant conditions, the successively expanding distance-based evacuation zone could have been the most reasonable approach.

However, after the onsite situation stabilized, the distance-based evacuation zone was not timely updated using radiation measurements. In particular, Japan had no predetermined deposition-dose criteria to advise the longer term emergency actions beyond the Emergency Planning Zone (EPZ), although international guidelines such as the 2007 ICRP [76] were available. This led to people staying in high-dose areas beyond the evacuation zone (such as Iitate village). The government adopted the ICRP criteria in April, and defined the Deliberate Evacuation Zone as areas with projected dose exceeding 20 mSv/yr. The actual relocation of some of the highly contaminated areas did not begin until May 15. In contrast, many areas with low dose rates, for example in Tamura and Kawauchi, remained as part of the evacuation zone for several years, which prevented the residents' return and exacerbated the relocation-related health consequences [77].

For the early phase of a nuclear emergency, international regulatory guidance requires that every NPP establishes an EPZ, i.e., an area with prepared arrangements to assure that prompt protective actions can be taken during a radiological accident [78]. Outside the EPZ, protective actions have to be developed at the time on a need basis, based on dose forecasts [79]. However, since Japan's radioactive release and dose prediction software, SPEEDI [80], failed to provide quantitative dose forecasts, early-stage protective actions beyond the EPZ were delayed and based on ad-hoc estimates. Following the accident, Japan revised their EPZ scope and expanded it to 30 km [81].

The current emergency response plans consider different evacuation strategies—for example, proposing a combination of evacuation and sheltering areas, depending on the expected path of the plume, rather than an isotropic/radial evacuation [82]. Nevertheless, any tailored evacuation scheme requires the availability of sufficient radiation data, as well as reliable real-time plume forecasts. Although there have been many developments in real-time radioactive atmospheric transport and dispersion (ATD) codes [83, 84], they still suffer from several limitations. To the authors' knowledge, even state-of-the-art codes (e.g. US LODI [85]) require source term information [85, 86], which can be difficult to quantify in real time during an accident. Moreover, many codes do not include sufficient uncertainty quantification capabilities, do not assimilate real-time radiation measurements into their predictions [87], and/or have very high computational costs that compromise their suitability for real-time emergency applications. These limitations were the reason behind the Japanese government's decision to avoid relying on ATD predictions in future emergencies [88].

**Emergency response planning and preparations.** Before the FDNPP accident, the scope of Probabilistic Risk Assessments (PRA) in Japan was limited and did not consider beyond-core-damage sequences (radioactive release or dose exposure) [70]. The probability of a severe accident was considered too low to occur, and severe accidents requiring evacuations were considered impossible [89]. Such perception had a significant influence on how severe accidents and emergency response plans were developed. Moreover, the national emergency management plan was formulated on the presumption that a natural disaster and a nuclear emergency would not take place concurrently. During the FDNPP accident, the interplay between the accident and the natural disasters aggravated the situation, which compromised many of the emergency plan's logistics (e.g., communication, power lines, and infrastructure).

Despite the fact that the evacuation of many patients was delayed, there were no radiation-related deaths in the evacuated patients [10]. In contrast, the disorganized and rushed evacuation of hospitals caused tens of fatalities, mainly because of the absence of hospital evacuation plans and medical arrangements prior to the accident. According to the National Diet of Japan [64], only the Imamura hospital (out of seven hospitals in the region) had an evacuation plan for a nuclear emergency.

In addition, the KI pill distribution was significantly delayed and ineffective, since the pills needed to be taken before, or at the time of, the exposure to radioactive iodine. The delay in implementing the KI measures resulted from the failure of the responsible authority to reach a consensus on the benefits (versus the side effects) of the pills [55]. To ensure timely intake, several countries pre-distribute the KI pills to the residents within the EPZ as part of their emergency planning [90]. Following the FDNPP accident, Japan instructed the pre-distribution of KI pills to the population in the vicinity of nuclear facilities [91].

At the same time, sheltering-in-place was not effectively executed during the FDNPP accident due to supply shortages. In particular, hospitals and healthcare facilities were not able to maintain their staffing and daily necessities, which led to increased mortality rates among the sheltering-in-place patients [92]. Although the IAEA recommends limiting sheltering in place to two days [93], the sheltering order was issued on March 15 and remained in place until April 22. Several municipalities had to voluntarily evacuate the sheltering residents, since they ran out of food and other critical supplies. This is primarily because the government did not have any pre-planned measures to cope with extended sheltering periods.

## Outlook and Resilience Design

The three key factors identified above enable us to propose the priority areas for further research and technical innovations for improving the resilience of the nuclear energy industry. To the authors' knowledge, these aspects have not been reported before.

**Meteorology-informed proactive containment venting.** Our analysis has highlighted the need to change the prevailing philosophy regarding containment venting. Specifically, during severe accidents, containment venting should be viewed as the primary means to protect the containment's integrity and prevent a significant release from "undesignated" pathways. SAMGs should facilitate early containment venting decisions, and the installation of FCVS can further reduce radionuclide release and environmental impacts. In addition, systems must include redundant vent lines that make use of passive concepts (e.g., hydrostatic pressure-based), fail-safe designs (e.g., fail-open isolation valves), and/or valves that can be operated manually outside or from a shielded area.

Additionally, wind forecasting—which we identify as a major driver for determining the accident impact—has been advancing rapidly in support of wind power generation and wildfire responses [94, 95]. We recommend that operators should have access to advanced meteorological forecasts to make use of any favorable conditions (e.g., wind blowing to unpopulated areas) when deciding to vent. The decision to vent should ultimately consider plant parameters (e.g., containment pressure and pressurization rate), the development of the situation on site (e.g., availability of power and radiological conditions), and the wind forecasts.

**Data-informed adaptive evacuation zones.** Protective actions and evacuation zones should be adaptive, informed by accident progressions, sensor datasets, and forecasts. The onsite accident progressions such as core status, containment venting plans, and expected source terms should inform the evacuation extent decisions. In fact, during the FDNPP accident, the evacuation zone extent was updated progressively following the onsite conditions from 2 km to 20 km. However, these decisions were not part of the emergency plan, which resulted in significant confusions.

Moreover, the evacuation zone and protective actions should be dynamically updated as radiation measurements and survey data become available, synthesizing airborne surveys and in situ monitoring posts [27, 96]. Machine learning (ML)-based methods have been developed recently for integrating various types of measurements, such as real-time radiation monitoring points and ground/airborne-based surveys [27, 97]. Such a framework has proven effective, particularly with the availability of mobile and rapidly deployable air and ground radiation monitoring platforms [96]. At the same time, in situ monitoring posts must have independent and reliable power sources in case of blackouts and should be placed at strategic locations. Although ML has been used for optimizing sensor networks for long-term radiation monitoring [98], further development is needed for efficiently placing monitoring posts to rapidly capture the plume extent in case of accidents. These developments can help in updating the contamination maps in real time, hence, supporting an adaptive evacuation-zone extent that would reduce the public radiation exposure, as well as allow residents back into some areas—earlier—if the dose rate is not high after the accident concludes. In particular, shrinking the evacuation zone is important to reduce unnecessary social and economic impacts as well as evacuation-induced health effects, given that prolonging the evacuation period tends to make the restoration process more difficult.

Reliable plume forecast, or even wind/weather forecasts, can guide evacuation directions and identify the priority areas for evacuation, particularly when the plume is expected to move beyond the pre-planned area. In addition, ML is rapidly being applied to weather forecasting to accelerate simulations through surrogate modeling, to assimilate real-time monitoring data, and to improve the prediction accuracy and uncertainty quantification. Specifically, combining physical models (such as plume transport models) and real-time datasets is an active area of research (such as physics-informed ML) for improving both forecasting and mapping [99-101]. These ML developments need to be integrated into the ATD codes for improving plume prediction accuracy and informing real-time emergency response.

**Comprehensive risk-informed emergency planning.** Off-site accident consequence assessments such as PRA level 3 (PRA-3) are critical for informing the EPZ extents and evacuation plans. The PRA-3 develops the potential accident scenarios, estimates the offsite radiological consequences, with uncertainty considered, and assesses what protective measures are effective, at which location and at what time. PRA-3 results can therefore guide the necessary emergency preparations, including the needed technological and human resources, as well as essential stockpiles (e.g., food, fuel, and medications). However, at the moment, PRA-3 is not a regulatory requirement in most countries [4, 102]; thus, we recommend mandating it as part of the licensing process to push its development and narrow down uncertainties. Moreover, recent risk studies have assessed that the major threat facing the current nuclear fleet are external natural hazards [103]. Therefore, the potential for another concurrent natural-nuclear compound event should be well reflected in PRA-3 assessments and emergency management plans [104].

Furthermore, the FDNPP accident suggested that the health consequences were not just radiation-induced but could also result from evacuation, long-term relocation, and the associated psychological and social effects. Therefore, emergency response frameworks should consider all risk dimensions, which should be supported by data-informed risk-benefit analysis of the different protective actions. However, while nuclear accidents are rare and the corresponding response data are limited, other disaster classes, such as wildfires, hurricanes, and floods, occur in higher frequency and trigger large-scale evacuations and prolonged relocation [105, 106]. Specifically, evacuation-related risks have been well documented in these disasters [107], and evacuation-dynamics models and strategies are rapidly developing in their respective research communities [32, 108, 109]. As such, we find it critically important for the scientific community to exchange relevant empirical data and strategies across different types of disasters. This can improve evacuation logistics and execution—including social behaviors and traffic—and enhance the resilience of nuclear power systems.

## Methods

**FDNPP description.** Units 1-3 had Mark I PCVs comprising a bulb-shaped DW made of 30 mm steel, backed by reinforced concrete, and connected to a torus-shaped wetwell containing the suppression pool beneath. During normal operation, the PCV pressure is maintained between 0.115-0.13 MPa. The absolute design pressure of the PCV is 0.52 MPa for Unit 1, and 0.48 MPa for Units 2 and 3 [35]. Furthermore, to prevent a premature venting, the PCV vent lines contained a rupture disk that was set to passively burst when the PCV pressure exceeds the PCV design pressure [28].

**Plant status before the accident**. When the earthquake occurred, Units 1, 2, and 3 were in normal operation with the RPV pressure at 7 MPa. Units 4, 5, and 6 were shut down for refueling and maintenance outage; hence those units were affected to a lesser extent. Unit 4 had its fuel offloaded to the spent-fuel pool, while Units 5 and 6 had their fuel assemblies in the core [35].

**PCV failure modes.** The "Containment Integrity Program" at the Sandia National Laboratory [110] defines two modes of PCV failure when the PCV exhibits beyond-design loads. The first is a functional failure, which is a progressive failure in which the PCV experiences beyond-design leakage rates associated with smooth and continuous pressure decrease. The second is a structural failure characterized by large global deformations and potential catastrophic rupture associated with a sudden loss of PCV pressure.

PCV failure depends on many parameters, such as the load distribution, duration, and temperature; accordingly, the fragility curve is usually modeled probabilistically as a function of "above-design excess pressure" and time [110, 111]. Therefore, it is not straightforward to predict exactly at which point of beyond-design the containment will fail, and whether it will fail functionally or structurally. This explains why the PCVs in Units 1 and 3 did not fail structurally as in Unit 2, even though they witnessed beyond-design pressures, and for quite a period of time in the case of Unit 1.

**Synthesis methodology.** We reconstruct the timeline of the FDNPP accident, reactor dynamics, radioactive release, contamination, and evacuation zone situation, as well as the health effects, by synthesizing information from different sources. First, the timeline of the in-plant accident progression and government protective actions and emergency response are developed based on multiple official investigation reports [28, 29, 43, 112]. The RPV and PCV pressure dynamics of each unit are developed using the publicly available datasets provided by TEPCO [113]. The assigned timing of the physical events (such as core damage, breach of the RPV, core material relocation, hydrogen explosion, and similar phenomena) are based on severe accident computer code calculations [42, 45, 51, 114], the physical analysis in Ishikawa [47], as well as the temporal pressure dynamics in the plant.

The source term information is retrieved from a very recent evaluation by Terada et al. [20], which is based on reverse-estimate calculations, coupling atmospheric transport and dispersion (ATD) simulations with monitoring data to find the optimal – matching – release rates. We also compare the major source term peaks with the actual dose rate measurements recorded by TEPCO's monitoring car at the FDNPP [52]. Furthermore, we map the observed peaks to the concurrent onsite physical events that we could identify and anticipate to be behind these release patterns. We would note that the source term contains significant uncertainties (mainly from the ATD model assumptions and simulation results), and there can be some discrepancies with the assigned physical events, i.e., time lag or lead, release duration, and magnitude. Wind information during the accident is retrieved from the meteorological maps provided by the Japanese National Institute of Informatics [115]. The dose rate maps of the Fukushima prefecture provided by the Prefectural Government (using real-time dosimeters) [116] supports the study of the contaminated regions, contribution of the main plumes, and time evolution of the evacuation zones.

Radiation exposure statistics are based on the UNSCEAR report [58] and the surveys in Nagataki et al. [59]. We use the data provided by the National Diet of Japan [64], Satoh et al. [63], Bromet [61], Yabe et al. [62], and Igarashi et al. [65] to examine the nonradiation health effects and establish the collateral damage caused by evacuations, family separations, social distress, induced mental health problems, and patients' relocations.

## Acknowledgements

The authors would like to thank the Japan Atomic Energy Agency scientists Hiroaki Terada and Haruyasu Nagai for their thorough review and constructive feedback especially when it comes to the source term estimation and its relation to onsite events. We would also like to thank TEPCO's engineers and scientists Tatsuro Kobayashi, Masui Hideki, and Shinya Mizokami for their help in resolving several inquiries we had regarding the accident development, data, and post-accident retrofits. Furthermore, we thank Prof. Jacopo Buongiorno (MIT) for his constructive feedback on the manuscript, as well as Prof. Horst-Michael Prasser (Prof. em. ETH Zurich) for his valuable insights on containment designs and future needs. Finally, the authors would like to thank Russell Prentice (Pacific Gas and Electric Company) for the valuable discussions on the current emergency response practices in the nuclear industry.

## Author contributions

A.A. and H.M.W. conceptualized the study. A.A. conducted the literature review, the analyses, and prepared the initial manuscript. A.A. and H.M.W. curated the analyses, interpreted the results, and revised the manuscript. G.S., R.G. and K.S. helped in the validation, writing–review and manuscript editing.

## Competing interests

The authors declare no competing interests.